\documentclass[prl,aps,twocolumn,superscriptaddress,showpacs]{revtex4}
\usepackage{graphicx,color}

\definecolor{yellowx}{rgb}{0.996,0.996,0.598}
\definecolor{bluex}{rgb}{0.0,0.0,0.5}
\definecolor{greenx}{rgb}{0.0,0.5,0.0}
\definecolor{redx}{rgb}{0.8,0.0,0.0}

\usepackage{graphicx}
\usepackage{amsmath,amssymb}

\begin{document}

 {\noindent \large \bf Supplementary Material to the Comment
on ``Photoluminescence Ring Formation in Coupled Quantum Wells:
Excitonic Versus Ambipolar Diffusion''}

\bigskip

\noindent A.\,L.~Ivanov$^1$, E.\,A.~Muljarov$^1$,
L.~Mouchliadis$^1$, and R.~Zimmermann$^2$

{\noindent \small $^1$ Department of Physics and Astronomy,
Cardiff University, Cardiff CF24 3AA, United Kingdom}

{\noindent \small $^2$ Institut f\"ur Physik der
Humboldt-Universit\"at zu Berlin, Berlin 12489, Germany}
\bigskip

\noindent Due to the lack of space in our one-page
Comment~\cite{Comment} on a Letter~\cite{Stern08}, below we extend
our arguments, concentrating mainly on (i) the validity of
Eq.\,(1) proposed in the Comment to include the screening effect
in exciton-exciton dipole-like interaction and (ii) the quantum
mass action law (QMAL) for ground-state indirect excitons screened
by free carriers.
 \bigskip

\noindent {\bf (i)} Equation (1) in Ref.\,\cite{Comment} implies a
quasi-equilibrium state for indirect excitons. Note, that for a
system to be in quasi-equilibrium and to have a constant chemical
potential, $\mu = \mbox{const}$, there is no need to require the
mean free path $l$ of particles to be much less than a
characteristic length scale $r_0$ of the potential of
particle-particle interaction. For example, the Thomas-Fermi
screening radius in metals $r_0\lesssim 1$\,nm, while usually the
low-temperature mean free path of electrons $l\gtrsim 100$\,nm.
This means that $l \gg r_0$, but the Thomas-Fermi formula, which
requires an equilibrium state for electrons with a well-defined
chemical potential $\mu$, is of course correct. In the famous book
by D. Pines \cite{Pines1963}, the derivation of the Thomas-Fermi
screening for electrons in metals is detailed with no reference to
the relationship between $r_0$ and $l$. On the the other hand, the
time needed for a system to relax to its (quasi-) equilibrium
state does depend on characteristic length scales. In the case of
indirect excitons, quasi-equilibrium establishes in $10 -
100$\,ps, i.e., in a time scale much shorter than characteristic
times of the drift-diffusion and optical decay processes.

Conceptually, our approach to the screening of exciton-exciton
interaction is similar to that of electron-electron interaction in
theory of metals. In metals, the Thomas-Fermi screening (the
random phase approximation, in a more technical way of saying)
changes the $1/r$ long-range electron potential to a local one,
$e^{-r/r_0}/r$. In a similar way, for indirect excitons the
thermal screening, Eq.\,(1) in our Comment~\cite{Comment},
effectively cuts the bare mid-range $1/r^3$ interaction.
\newpage

\noindent {\bf (ii)} Generally, the QMAL should include
self-consistently the change of the exciton state, due to
screening and phase space filling associated with free $e$-$h$
pairs \cite{Snoke2008}. In inset (b) of Fig.\,1 of the
Comment~\cite{Comment}, we show how the total number of $e$-$h$
pairs is distributed among the bound (exciton) and unbound states,
according to the QMAL and {\it taking into account screening of
indirect excitons by free carriers}. However, for the parameters
relevant to the experiments \cite{Stern08}, the screening effect
practically does not change the result. In the present Fig.\,1 we
plot the density $n$ of indirect excitons, which is calculated
with the QMAL for the constant binding energy $\epsilon_{\rm x}$
with no screening effect due to free carriers (red solid lines),
versus that evaluated by using the QMAL with the screening effect
which changes the exciton binding energy $\epsilon_{\rm x}$ (blue
dotted lines). The screening effect is included according to
Eq.\,(17) of Ref.\,\cite{Snoke2008}. As it is clearly seen from
Fig.\,1, the difference is rather minor indeed, due to $n \gg
n_{\rm e} = n_{\rm h}$.
\begin{figure}[t]
\begin{center}
\includegraphics*[width=7.0cm,angle=0]{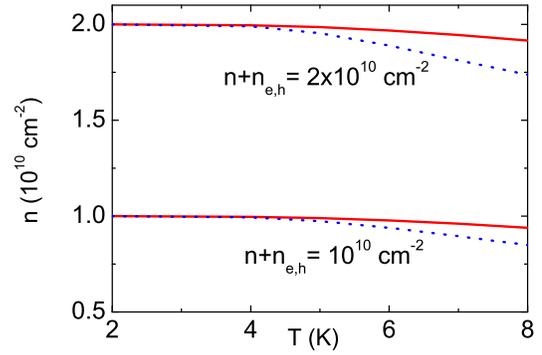}
\caption{Exciton concentration calculated with the QMAL for the
constant binding energy $\epsilon_{\rm x} = 3.5$\,meV with no
screening effect due to free carriers (red solid lines) against
that evaluated by using the QMAL with the screening effect (blue
dotted lines). The total concentration of excitons and free
$e$-$h$ pairs is 1 and $2\times 10^{10}$cm$^{-2}$.}
\end{center}
\end{figure}

\end{document}